\documentclass[10pt, conference, compsocconf]{IEEEtran}

\usepackage{times}
\usepackage{amsmath}
\usepackage{stmaryrd}
\usepackage{amssymb}
\usepackage{graphicx}
\usepackage[T1]{fontenc}

\usepackage[ruled]{algorithm2e}

\newcommand{\tr}{{\rm tr}}

\def\>{\ensuremath{\rangle}}
\def\<{\ensuremath{\langle}}

\def\h{\ensuremath{\mathcal{H}}}

\newtheorem{defn}{Definition}[section]

\begin{document}
\title{Model Checking Quantum Systems --- A Survey}

\author{\IEEEauthorblockN{Mingsheng Ying and Yuan Feng}
\IEEEauthorblockA{Centre for Quantum Software and Information, 
University of Technology Sydney,
Australia\\ State Key Laboratory of Computer Science, Institute of Software, Chinese Academy of Sciences, China\\ Department  of Computer Science and Technology, Tsinghua University, China\\
Email: Mingsheng.Ying@uts.edu.au; Yuan.Feng@uts.edu.au}
}

\maketitle

\begin{abstract}
This article discusses the essential difficulties in developing model-checking techniques for quantum systems that are never present in model checking classical systems. It further reviews some early researches on checking quantum communication protocols as well as a new line of researches pursued by the authors and their collaborators on checking general quantum systems, applicable to both physical systems and quantum programs.    
\end{abstract}

\IEEEpeerreviewmaketitle

\section{Introduction}

We are currently in the midst of a second quantum revolution: \textit{transition from quantum
theory to quantum engineering}~\cite{DM03}. The aim of quantum theory is to find
fundamental rules that govern the physical systems already existing
in nature. Instead, quantum engineering intends to design and
implement new systems (machines, devices, etc) that do not exist
before to accomplish some desirable tasks, based on quantum theory.
Active areas of quantum engineering includes quantum computing, quantum cryptography, quantum communication, quantum sensing, quantum simulation, quantum metrology and quantum imaging.

Experiences in today's engineering indicate that it is not
guaranteed that a human designer completely understands the
behaviours of the systems she/he designed, and a bug in her/his
design may cause some serious problems and even disasters. So,
correctness, safety and reliability of complex engineering systems
have attracted wide attention and have been systematically studied in
various engineering fields. In particular, in the last four decades, computer scientists have developed various verification techniques for the correctness of both hardware and software as well as the security of communication protocols. 

\subsection{Second Quantum Revolution Requires New Verification Techniques} 

As is well-known, human intuition is
much better adapted to the classical world than the quantum world.
This implies that human engineers will commit many more faults in
designing and implementing complex quantum systems such as quantum computer hardware and software and quantum communication protocols. Thus,
correctness, safety and reliability problems will be even more
critical in quantum engineering than in today's engineering. However, due to the essential differences between the classical and quantum worlds, verification techniques developed for classical engineering systems cannot be directly used to quantum systems. Novel verification techniques will be indispensable for the coming era of quantum engineering and quantum technology \cite{Zo12}.  

\subsection{Model Checking Techniques for Classical Systems}

Model-checking is an effective automated technique that checks whether a desired  
property is satisfied by a system, e.g. a computing or communication system. The properties that are checked are usually specified in a logic, in particular, temporal logic; typical properties are deadlock freedom, invariants, safety, request-response properties. The systems under checking are mathematically modelled as e.g. (finite-state) automata, transition systems, Markov chains and Markov decision processes \cite{Clarke01, BK08}. 

Model-checking has become one of the dominant techniques for verification of computer (hardware and software) systems 30 years after its inception. Many industrial-strength
systems have been verified by employing model-checking techniques.
Recently, it has also successfully been used in systems biology; see~\cite{HKNPT06} for example. 

With quantum engineering and quantum technology being emerging, a question then naturally arises: \textit{is it possible and how to
use model-checking techniques to verify correctness and safety of quantum engineering systems}?

\subsection{Difficulty in Model Checking Quantum Systems}\label{probs}  
Unfortunately, it seems that the current model-checking
techniques cannot be directly applied to quantum systems because of
some essential differences between the classical world and the quantum world. To develop model-checking techniques for quantum systems, the following three problems must be systematically addressed: 
\begin{itemize}\item \textbf{\textit{System modelling and property specification}}: 
The classical system modelling method cannot be used to describe the behaviours of quantum systems, and the classical specification language is not suited to formalise the properties of quantum systems to be checked.      
So, we need to
carefully and clearly define a conceptual framework in which we can
properly reason about quantum systems, including \textit{formal models of
quantum systems} and \textit{formal description of temporal properties of
quantum systems}.

\item \textbf{\textit{Quantum measurements}}: Model-checking is usually applied to check long-term behaviours of the systems. But to check whether a quantum system satisfies a certain property at a time point, one has to perform a quantum measurement on the system, which can change the state of the system. This makes studies of the long-term behaviours of quantum systems much harder than that of classical systems \cite{Gr96, BP02, Br02}.

\item \textbf{\textit{Algorithms}}: The state spaces of the classical systems that model-checking algorithms can be applied to are usually finite or countably infinite. However, the state spaces of quantum systems are inherently continuous even when they are finite-dimensional. In order to develop algorithms for model-checking quantum systems, we have to exploit some deep mathematical properties of the systems so that it suffices to examine only a finite number of (or at most countably infinitely many) representative elements, e.g. those in an orthonormal basis, of their state spaces. Also, a linear algebraic structure always resides in the state space of a quantum system. So, an algorithm checking a quantum system should be carefully developed so that the linear algebraic structure will not be broken.
\end{itemize} 

\section{Early Research on Model Checking of Quantum Systems}\label{early}

Despite the difficulties discussed in the previous section, quite a few model-checking techniques for quantum systems have been developed in the last 10 years. The earliest work mainly targeted checking quantum communication protocols:\begin{itemize}\item  
Taking the probabilism arising from quantum measurements into account, \cite{GNP06} used the probabilistic model-checker PRISM~\cite{KNP04} to verify the correctness of quantum protocols, including superdense coding, quantum teleportation and quantum error correction.

\item A branching-time temporal extension (called quantum computation tree logic or QCTL for short) of exogenous quantum propositional logic \cite{MS06} was introduced and then the model-checking problem for this logic was studied in \cite{BCMS07, BCM08}, with verification of the correctness of quantum key distribution BB84 \cite{BB84} as an application.

\item A linear temporal extension QLTL of exogenous quantum propositional logic \cite{MS06} was then defined and the corresponding model-checking problem was investigated in \cite{Mat10}. 

\item Model-checking techniques were developed in \cite{Da11, DG12} for quantum communication protocols modelled in process algebra CQP (Communicating Quantum Processes) \cite{GN05}. The checked properties are specified by the quantum computation tree logic QCTL defined in \cite{BCM08}. 

\item A model-checker for quantum communication protocols was also developed in \cite{GNP08, Pa08, GNP10}, where the checked properties are specified by QCTL \cite{BCM08} too, but only the protocols that can be modelled as quantum circuits expressible in the stabiliser formalism \cite{Go97} were considered. In \cite{AGN13, AGN14}, this technique was extended beyond stabiliser states and used to check equivalence of quantum protocols. 
\end{itemize}

\section{Model Checking Quantum Automata}

A research line pursued by the authors and their collaborators is to develop model-checking techniques that can be used not only for quantum communication protocols but also for general quantum systems, including physical systems and quantum programs.  

Quantum automata were adopted in \cite{TOCL13, Concur'14} as the model of the systems:

\begin{defn}[Quantum automata \cite{KW97, Mo00}]\label{q-auto} A quantum automaton is a $4$-tuple $\mathcal{A}=(\mathcal{H},\mathit{Act}, \{U_\alpha:\alpha\in\mathit{Act}\},\mathcal{H}_0)$,
where:\begin{enumerate}
\item $\mathcal{H}$ is a finite-dimensional Hilbert space, called the state space;
\item $\mathit{Act}$ is a finite set of action names;
\item for each action name $\alpha\in\mathit{Act}$, $U_\alpha$ is a unitary operator on $\mathcal{H}$; \item $\mathcal{H}_0\subseteq\mathcal{H}$ is the subspace of initial states.\end{enumerate}
\end{defn}

A quantum automaton behaves as follows: it starts from some initial state in $\mathcal{H}_0$, and at each step it performs a unitary transformation $U_\alpha$ for some $\alpha\in\mathit{Act}$.  
An algorithm for checking certain linear-time properties (e.g. invariants and safety properties) was proposed in \cite{TOCL13}, where following Birkhoff-von Neumann quantum logic \cite{BN36}, closed subspaces of the state Hilbert space are used  as the atomic propositions about the state of system, and the checked linear-time properties are defined as infinite sequences of sets of atomic propositions. Furthermore, decidability or undecidability of several reachability problems (namely, eventually reachable, globally reachable, ultimately forever reachable, and infinitely often reachable) for quantum automata were established in \cite{Concur'14}.

\section{Model Checking Quantum Markov Chains}

The model-checking problem for a larger class of quantum systems than quantum automata, namely quantum Markov chains and quantum Markov decision processes was studied in a series of papers by the authors and their collaborators \cite{YY12a, Concur'13, IC'18, JCSS'18}.  

Continuous-time quantum Markov processes have been intensively studied in mathematical physics, and discrete-time quantum Markov chains were introduced in \cite{YY12} as a semantic model for the purpose of termination analysis of quantum programs.

\begin{defn}[Quantum Markov chains \cite{YY12}] A quantum Markov chain is a triple $(\mathcal{H},\mathcal{E},\mathcal{H}_0)$, where $\mathcal{H}$ and $\mathcal{H}_0$ are the same as in Definition \ref{q-auto}, and $\mathcal{E}$ is a super-operator on $\mathcal{H}$.
\end{defn}

A quantum Markov chain starts in an initial state in $\mathcal{H}_0$, and at each step it performs (the same) quantum operation modelled by the super-operator $\mathcal{E}$. Note that the (discrete-time) dynamics of closed quantum systems are usually depicted by unitary operators, and the behaviours of open quantum systems are described by super-operators (see \cite{NC00}, Section 8.2). Obviously, the notion of quantum automata can be generalised by replacing unitary operators $U_\alpha$ $(\alpha\in\mathit{Act})$ in Definition \ref{q-auto} by super-operators $\mathcal{E}_\alpha$ $(\alpha\in\mathit{Act})$. Furthermore, quantum Markov decision processes can be defined by introducing decision strategies into such generalised quantum automata \cite{Barry14, IC'18}.  

Several algorithms for checking reachability of quantum Markov chains and quantum Markov decision processes were developed in \cite{Concur'13, IC'18}. As in checking classical Markov chains and Markov decision processes, graph reachability is a key to these algorithms. However, classical graph theory is not suited to our purpose; instead a new theory of quantum graphs (i.e. graphs in a Hilbert space with adjacency relation induced by a super-operator) was developed, and in particular, an algorithm for the BSCC (bottom strongly connected components) decomposition of the state Hilbert spaces was found in \cite{Concur'13}. Another decomposition technique, namely periodic decomposition, for quantum Markov chains was recently proposed in \cite{JCSS'18}. 

\section{Model Checking Super-Operator-Valued Markov chains}

The notion of super-operator-valued Markov chain is introduced in \cite{JCSS'13} as a higher-level model of quantum programs and quantum cryptographic protocols. A similar notion was proposed in \cite{Gudder} for a different purpose.

\begin{defn}[Super-operator-valued Markov chains \cite{JCSS'13}]
A labelled super-operator-valued Markov chain over a set $AP$ of predefined atomic propositions is a $5$-tuple $(S, s_0, \mathcal{H}, Q, L)$, where:
\begin{enumerate}
	\item $S$ is a finite set of \emph{classical states} with $s_0\in S$ being the initial state;
	\item $\mathcal{H}$ is a finite-dimensional Hilbert space, called the quantum state space;
	\item  $Q : S\times S\rightarrow \mathcal{SO}_\h$ is a transition super-operator function, where $\mathcal{SO}_\h$ denotes the set of trace-nonincreasing super-operators on $\mathcal{H}$, and for each $s\in S$, $\sum_{t \in S} Q(s,t)$ is trace-preserving; and
	\item   $L: S \to 2^{AP}$ is a labelling function.
\end{enumerate}
\end{defn} 

A super-operator-valued Markov chain has two state spaces, a classical one and a quantum one, which are connected through the transition super-operator function. It behaves in a similar manner as classical Markov chains. It starts from the classical initial state $s_0$ but with the quantum initial state unspecified (it can be taken arbitrarily). Then at each step, given the current classical state $s$ and quantum state $\rho$, it proceeds to classical state $t$ with probability $\tr[Q(s,t)(\rho)]$, and the accompanied quantum state evolves into $Q(s,t)(\rho)/\tr[Q(s,t)(\rho)]$ provided that $\tr[Q(s,t)(\rho)]\neq 0$. The normalisation requirement that  $\sum_{t \in S} Q(s,t)$ is trace-preserving guarantees that the probabilities of going from $s$ to some classical state sum up to 1.

As the atomic propositions are taken to be classical (they apply only to classical states), this Markov chain model is suitable for verification of quantum systems against classical properties, such as running time, termination, reachability, etc.
One distinct feature of this model, however, for verification purpose, is that it provides a way to check \emph{once-for-all} in that once a property is checked to hold, it holds for all initial quantum states.
For example, for the reachability problem, the model checking algorithm essentially calculates a positive operator $\Pi$, accounting for all (classical) paths satisfying the concerned property.
Then the reachability \emph{probability} when the Markov chain is started in the initial quantum state $\rho$ is simply $\tr(\Pi \rho)$.

A corresponding computation tree logic (CTL) for super-operator-valued Markov chains was defined, and algorithms for checking such properties were developed in~\cite{JCSS'13}. A tool implementation of these algorithms has been provided~\cite{FengHTZ15} based on the probabilistic model checker~\textsc{IscasMC}~\cite{HahnLSTZ14}. Algorithms for model checking $\omega$-regular properties, a very general class of properties subsuming those expressible by LTL formulae, against super-operator-valued Markov chains were proposed in~\cite{concur17}. This allows to express and analyse a wide range of relevant properties, such as repeated reachability, reachability in a restricted order, nested Until properties, or conjunctions of such properties. Furthermore, the reachability problem of a recursive extension of super-operator-valued Markov chains was studied in \cite{MFCS'13}, with the application of analysing quantum programs with procedure calls.

\section{Conclusion}

As reviewed in previous sections, several theoretical frameworks and algorithms of quantum model-checking have been developed. But certainly, quantum model-checking is still at a very early stage of its development; in particular, its applications are only at the level of toy examples. We envisage that in the future, quantum model-checking techniques can be applied to the following areas:
\begin{enumerate}\item \textit{Checking physical systems}: Physicists already considered the algorithmic checking problem of certain properties of quantum systems, for example, quantum measurement occurrence \cite{EMG12} and reachability of quantum states \cite{SSL02}. Quantum model-checking can offer a systematic view of this line of research. 
\item \textit{Verification of quantum circuits}: Verification of circuits has been one of the major application areas of classical model-checking. But model-checking applied to verification of quantum circuits is an area to be systematically exploited.  
\item \textit{Analysis and verification of quantum programs}: Another important application area of classical model-checking is analysis and verification of programs. Several techniques for analysis and verification of quantum programs have been reported in the last few years \cite{Ali, Yin11, Ying16, POPL'17, POPL'18}. However, model-checking techniques specifically designed for quantum programs are still missing.  
\item \textit{Verification of security of quantum communication protocols}: Applications of model-checking mentioned in Section \ref{early} focus on verification of correctness of quantum communication protocols. But verification of the security of quantum protocols is much more difficult, and model-checking applied to it is an interesting topic for future research. 
\end{enumerate}  Finally, a crucial step toward real-world applications of model-checking would be building efficient automatic tools.


\begin{thebibliography}{99}

\bibitem{AT12} C. Altafini and F. Ticozzi, Modeling and control of quantum systems: An introduction, \textit{IEEE Transactions on Automatic Control}, 57(2012)1898.

\bibitem{AGN13} E. Ardeshir-Larijani, S. J. Gay and R. Nagarajan, Equivalence checking of quantum protocols, in: \textit{Proceedings of the 19th International Conference on Tools and Algorithms for the Construction and Analysis of Systems (TACAS)}, Springer LNCS 7795, 2013, pp. 478-492.

\bibitem{AGN14} E. Ardeshir-Larijani, S. J. Gay and R. Nagarajan, 
Verification of concurrent quantum protocols by equivalence checking, \textit{Proceedings of the 20th International Conference on Tools and Algorithms for the Construction and Analysis of Systems (TACAS)}, Springer LNCS 8413, 2014, pp. 500-514. 

\bibitem{BK08}{\sc Baier, C} {\sc and} {\sc Katoen, J. -P.} 2008. \newblock {\em Principles of Model Checking}. 
MIT Press, Cambridge, Massachusetts.

\bibitem{BCMS07}{\sc Baltazar, P.} {\sc Chadha, R.} {\sc Mateus, P.} {\sc and} {\sc  Sernadas, A.} 2007. \newblock Towards model-checking quantum
security protocols. \newblock In: P. Dini et al. (eds.), {\em Proceedings of the 1st International Conference on Quantum, Nano, and Micro Technologies (ICQNM'07)}. IEEE Press, 14.

\bibitem{BCM08}{\sc Baltazar, P.} {\sc Chadha, R.} {\sc and} {\sc Mateus, P.} 2008. \newblock Quantum
computation tree logic - model checking and complete calculus. 
\newblock {\em International Journal of Quantum Information\/}~{\em 6}, 219--236.

\bibitem{Barry14} J. Barry, d. T. Barry and S. Aaronson, Quantum partially observable Markov decision processes, \textit{Physical Review A} 90(2014), art. no. 032311. 

\bibitem{BB84}{\sc Bennett, C. H.} {\sc and} {\sc Brassard, G.} 1984. \newblock Quantum cryptography: public key distribution and coin tossing. \newblock In: {\em Proceedings of International Conference on Computers, Systems and Signal Processing\/}. 175--179.

\bibitem{BN36} G. Birkhoff and J. von Neumann, The Logic of Quantum Mechanics, \textit{Annals of Mathematics}, 37(1936)823-843.

\bibitem{BP02}{\sc Breuer, H. -P.} {\sc and} {\sc Petruccione, F.} 2002. \newblock {\em The Theory of Open Quantum Systems}. Oxford University Press, Oxford. 

\bibitem{Br02}{\sc Brun, T.} 2002. \newblock A simple model of quantum trajectories. \newblock {\em American Journal of Physics\/}~{\em 70}, 719--737.

\bibitem{Zo12}{\sc Cirac, J. I.} {\sc and} {\sc Zoller, P.} 2012. \newblock Goals and opportunities in quantum simulation. \newblock {\em Nature Physics\/}~{\em 8}, 264--266.

\bibitem{Clarke01}{\sc Clarke, E. M.} {\sc Grumberg, O.} {\sc and} {\sc Peled, D. A.} 2011. \newblock {\em Model Checking}. MIT Press, Cambridge, Massachusetts.

\bibitem{Da11}{\sc Davidson, T.} 2011. \newblock {\em Formal Verification Techniques using Quantum Process Calculus}, PhD thesis, University of Warwick.

\bibitem{DG12}{\sc Davidson, T.} {\sc Gay, S. J.} {\sc Mlnarik, H.} {\sc Nagarajan, R.} {\sc and} {\sc Papanikolaou, N.} 2012. \newblock Model checking for Communicating Quantum Processes. 
\newblock {\em International Journal of Unconventional Computing\/}~{\em 8}, 73--98.

\bibitem{DM03}{\sc Dowling, J. P.} {\sc and} {\sc Milburn, G. J.} 2003. \newblock Quantum technology: the second quantum revolution. \newblock {\em Philosophical Transactions of the Royal Society London A\/}~{\em 361}, 1655--1674.

\bibitem{EMG12} J. Eisert, M. P. M\"{u}ller and C. Gogolin, Quantum measurement occurrence is undecidable, \textit{Physcal Review Letters}, 108(2012)260501.

\bibitem{JCSS'13} Y. Feng, N. K. Yu and M. S. Ying, Model checking quantum Markov chains, \textit{Journal of Computer and System Sciences}, 79(2013) 1181-1198. 

\bibitem{FengHTZ15} Y. Feng, E. M. Hahn, A. Turrini, and L. Zhang.
\newblock {QPMC}: A model checker for quantum programs and protocols.
\newblock In {\em FM'15}, volume 9109 of {\em Lecture Notes in Computer
  Science}, pages 265--272. Springer, 2015.

\bibitem{concur17} Y. Feng, E. M. Hahn, A. Turrini, and S. Ying.
Model Checking Omega-regular Properties for Quantum Markov Chains.
In: \textit{Proceedings of the Concur}, 2017, pp. 35:1--35:16.

\bibitem{MFCS'13} Y. Feng, N. K. Yu and M. S. Ying, Reachability analysis of recursive quantum Markov chains, \textit{Proceedings of MFCS}, 2013, pp. 385-396.

\bibitem{GN05} S. J. Gay and R. Nagarajan, Communicating Quantum Processes, in: \textit{Proceedings of the 32nd ACM Symposium on Principles of Programming Languages (POPL)}, 2005, pp. 145-157. 

\bibitem{GNP06}{\sc Gay, S. J.} {\sc Nagarajan, R.} {\sc and} {\sc Papanikolaou, N.} 2005.
\newblock Probabilistic model-checking of quantum protocols \newblock {\em arXiv:quant-ph/0504007}.

\bibitem{GNP08}{\sc Gay, S. J.} {\sc Nagarajan, R.} {\sc and} {\sc Panaikolaou, N.} 2008. \newblock QMC: a
model checker for quantum systems. \newblock In: {\em Proceedings of the 20th International Conference 
on Automated Verification (CAV'08)}, Lecture Notes in Computer Science 5123, Springer, 
543--547.

\bibitem{GNP10}{\sc Gay, S. J.} {\sc Nagarajan, R.} {\sc and} {\sc Panaikolaou, N.} 2010. \newblock Specification and verification of quantum protocols. \newblock In: I. Mackie
and S. Gay (eds.), {\em Semantic Techniques in Quantum
Computation}, Cambridge University Press, 414--472.


\bibitem{Go97}{\sc Gottesman, D.} 1997. {\em Stablizer Codes and Quantum Error
Correction}, Ph.D. thesis, California Institute of Technology.

\bibitem{Gr96}{\sc Griffiths, R. B.} 1996. \newblock Consistent histories and quantum reasoning. \newblock {\em Physical Review A\/}~{\em 54}, 2759--2774.

\bibitem{JCSS'18} J. Guan, Y. Feng and M. S. Ying 2018. Decomposition of quantum Markov chains and its applications, 
\textit{Journal of Computer and System Sciences}, 95: 55-68. 

\bibitem{Gudder} S. Gudder, Quantum Markov chains, \textit{Journal of Mathematical Physics}, 49(2008) art. no. 072105.

\bibitem{HahnLSTZ14}
E. M. Hahn, Y. Li, S. Schewe, A. Turrini, and L. Zhang.
\newblock {IscasMC}: A web-based probabilistic model checker.
\newblock In {\em FM'14}, volume 8442 of {\em Lecture Notes in Computer
  Science}, pages 312--317. Springer, 2014.
  
\bibitem{HKNPT06}{\sc Heath, J.} {\sc Kwiatkowska, M.} {\sc Norman, G.} {\sc Parker, D.} {\sc and} {\sc Tymchyshyn, O.} 2006. \newblock Probabilistic
model checking of complex biological pathways. \newblock In: C. Priami (ed.) {\em Proceedings of CMSB}, Lecture Notes in Computer Science 4210, Springer, 32--47.

\bibitem{Ali} A. JavadiAbhari, S. Patil, D. Kudrow, J. Heckey, A. Lvov, F. T. Chong and M. Martonosi, 
ScaffCC: Scalable compilation and analysis of quantum programs, \textit{Parallel Computing}, 45(2015)2-17.

\bibitem{KW97}{\sc Kondacs, A.} {\sc and} {\sc Watrous, J.} 1997. \newblock On the power of quantum
finite state automata. \newblock In: {\em Proc. 38th Symposium on Foundation of
Computer Science (FOCS'97)}, 66--75.

\bibitem{KNP04}{\sc Kwiatkowska, M.} {\sc Norman, G.} {\sc and} {\sc Parker, P.} 2004. \newblock Probabilistic symbolic model-checking with PRISM: a hybrid approach. \newblock {\em International Journal on Software Tools for Technology Transfer\/}~{\em 6}, 128--142.

\bibitem{Concur'14} Y. J. Li and M. S. Ying, (Un)decidable problems about reachability of quantum systems, In: \textit{Proceedings of the Concur}, 2014, Springer, pp. 482-496. 

\bibitem{POPL'18} Y. J. Li and M. S. Ying, Algorithmic analysis of termination problems for quantum programs, In: \textit{Proceedings of POPL}, 2018, pp. 35:1-35:29. 

\bibitem{MS06}{\sc Mateus, P.} {\sc and} {\sc Sernadas, A.} 2006. \newblock Weakly complete axiomatisation of exogenous quantum propositional logic. \newblock {\em Information and Computation\/}~{\em 204}, 771--794.

\bibitem{Mat10} P. Mateus, J. Ramos, A. Sernadas and C. Sernadas, Temporal logics for reasoning about quantum systems. \newblock In: I. Mackie
and S. Gay (eds.), {\em Semantic Techniques in Quantum
Computation}, Cambridge University Press, 389--413.

\bibitem{Mo00} C. Moore and J. Crutchfield, Quantum automata and quantum grammars, \textit{Theoretical Computer Science}, 237(2000) 275-306.

\bibitem{NC00}{\sc Nielsen, M. A.} {\sc and} {\sc Chuang, I. L.} 2000. 
\newblock {\em Quantum Computation and Quantum Information}, Cambridge
University Press.

\bibitem{Pa08}{\sc Papanikolaou, N. K. } 2008. \newblock {\em Model Checking Quantum
Protocols}, PhD Thesis, Department of Computer Science, University
of Warwick.

\bibitem{SSL02} S. G. Schirmer, A. I. Solomon and J. V. Leahy, Criteria for reachability of quantum states, \textit{Journal of Physics A: Mathematical and General}, 35(2002)8551-8562.

\bibitem{Yin11} M. S. Ying, Floyd-Hoare logic for quantum programs, \textit{ACM Transactions on Programming Languages and Systems}, (2011) art. no. 19.

\bibitem{Ying16} M. S. Ying, \textit{Foundations of Quantum Programming}, Morgan Kaufmann, 2016. 

\bibitem{TOCL13} M. S. Ying, Y. J. Li, N. K. Yu and Y. Feng, Model-checking linear-time properties of quantum systems, \textit{ACM Transactions on Computational Logic}, 15(2014), art. no. 22.

\bibitem{POPL'17} M. S. Ying, S. G. Ying and X. D. Wu, Invariants of quantum programs: characterisations and generation, \textit{Proceedings of POPL},  2017, pp. 818-832. 

\bibitem{YY12}{\sc Ying, M. S.} {\sc Yu, N. K.} {\sc Feng, Y.} {\sc and} {\sc Duan, R. Y.} 2013. \newblock Verification of quantum programs. \newblock {\em Science of Computer Programming\/}~{\em 78}, 679--1700.

\bibitem{Concur'13} S. G. Ying, Y. Feng, N. K. Yu and M. S. Ying, Reachability probabilities of quantum Markov chains, \textit{Proceedings of Concur'13}, Springer, pp. 334-348.

\bibitem{IC'18} S. G. Ying and M. S. Ying, Reachability analysis of quantum Markov decision processes, \textit{Information and Computation} 2018.  

\bibitem{YY12a}{\sc Yu, N. K.} {\sc and} {\sc Ying, M. S.} 2012.\newblock Reachability and termination analysis of concurrent quantum programs. \newblock In: {\em Proceedings of 
the 23rd International Conference on Concurrency Theory (CONCUR 2012)}. Lecture Notes in Computer Science 7454, Springer, 69--83.
  
\end{thebibliography}
\end{document}